\documentclass[twocolumn,prb,showpacs]{revtex4}
\usepackage{amsmath}
\usepackage{amssymb}
\usepackage{graphicx}
\usepackage{bm}
\usepackage{epstopdf}  
\usepackage{psfrag}
\usepackage{epsfig}
\usepackage{subfigure}
\usepackage{mathtools}
\usepackage{color}

\newcommand{\ve}[1]{\mathbf{#1}}
\newcommand{\te}[1]{\bar{\bar{#1}}}

\begin{document}

\title{Inverse Prism based on Temporal Discontinuity and Spatial Dispersion}

\author{Alireza Akbarzadeh}
\email{alireza.akbarzadeh@polymtl.ca}
\author{Nima Chamanara}
\author{Christophe Caloz}
\affiliation{Polytechnique Montr\'{e}al, Montr\'{e}al, QC H3T 1J4, Canada }

\date{\today} 

\begin{abstract}
We introduce the concept of the inverse prism as the dual of the conventional prism and deduce from this duality an implementation of it based on temporal discontinuity and spatial dispersion provided by anisotropy. Moreover, we show that this inverse prism exhibits the following three unique properties: chromatic refraction birefringence, ordinary-monochromatic and extraordinary-polychromatic temporal refraction, and linear-to-Lissajous polarization transformation.
\end{abstract}



\maketitle

\begin{figure}[!]
  \centering
  \subfigure{
  \psfrag{z}[c][c][1.0]{(a)}
  \psfrag{a}[c][c][1.0]{$z$}
  \psfrag{b}[c][c][1.0]{$x$}
  \psfrag{c}[c][c][1.0]{$z=z_{0}(x)$}
  \psfrag{d}[c][c][1.0]{$z<z_{0}(x)$}
  \psfrag{e}[c][c][1.0]{$z>z_{0}(x)$}
  \psfrag{f}[c][c][1.0]{$n=n_{1}$}
  \psfrag{g}[c][c][1.0]{$n=n_{2}(\omega)$}
  \includegraphics[width=0.45\columnwidth]{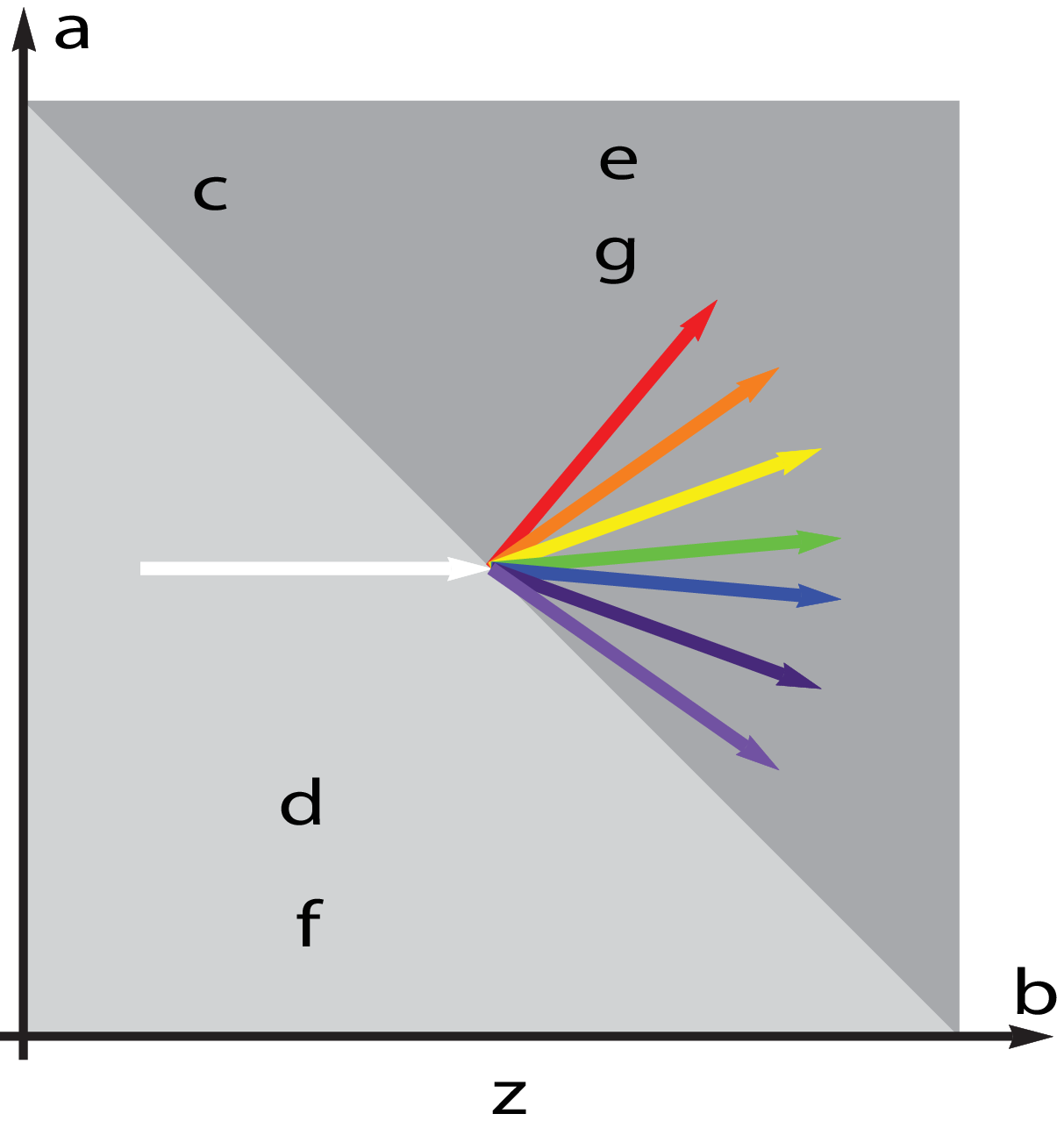}
  }
  \subfigure{
  \psfrag{z}[c][c][1.0]{(c)}
  \psfrag{a}[c][c][1.0]{$z$}
  \psfrag{b}[c][c][1.0]{$x$}
  \psfrag{c}[c][c][1.0]{$t=t_{0}$}
  \psfrag{d}[c][c][1.0]{$t<t_{0}$}
  \psfrag{e}[c][c][1.0]{$t>t_{0}$}
  \psfrag{f}[c][c][1.0]{$n=n_{1}$}
  \psfrag{g}[c][c][1.0]{$n=n_{2}(\ve{k})$}
  \includegraphics[width=0.45\columnwidth]{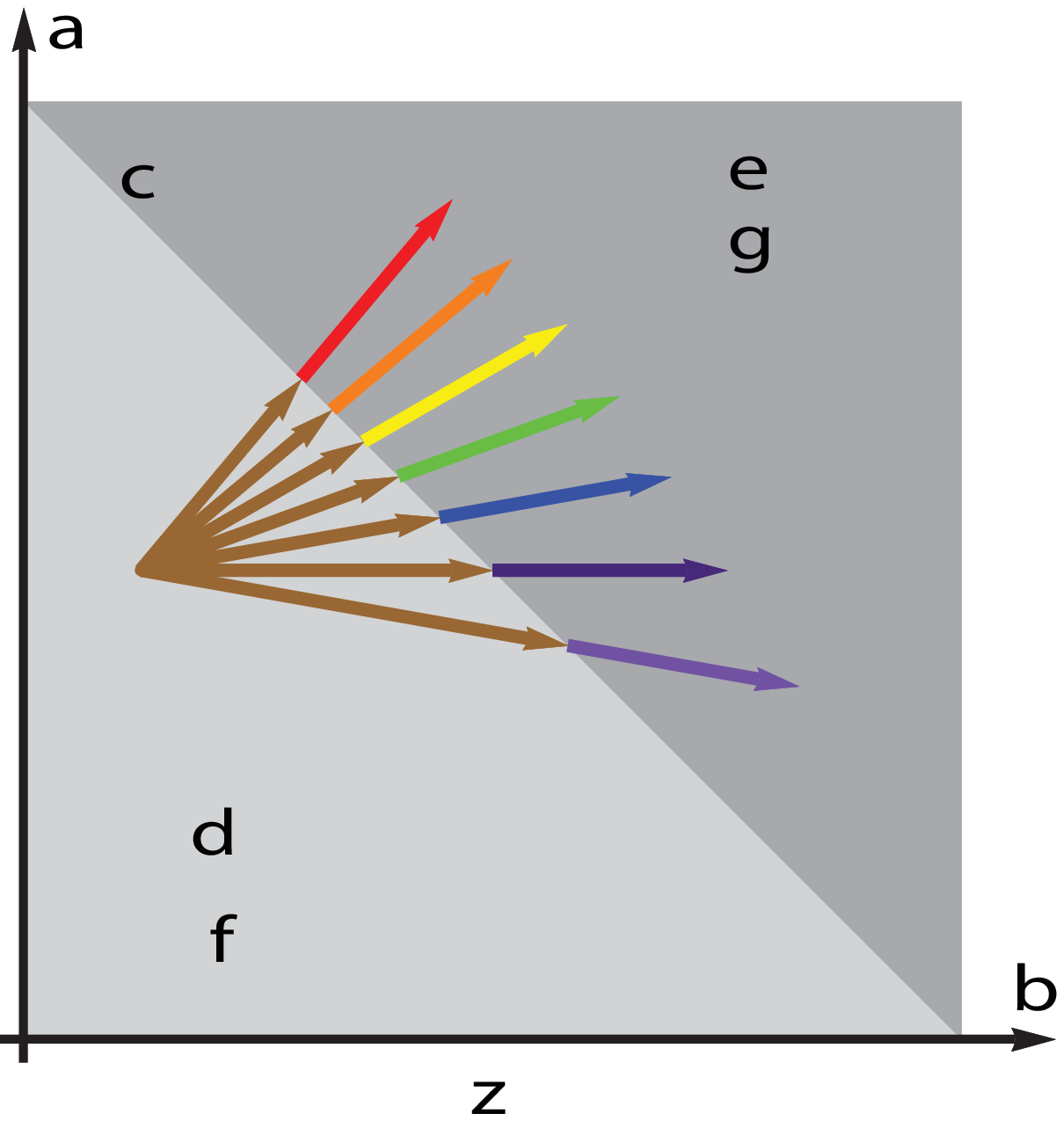}
  }
  \\
  \subfigure{
  \psfrag{z}[c][c][1.0]{(b)}
  \psfrag{a}[c][c][1.0]{$\omega_{a}$}
  \psfrag{b}[c][c][1.0]{$\omega_{b}$}
  \psfrag{c}[c][c][1.0]{$\omega_{c}$}
  \psfrag{d}[c][c][1.0]{$\omega_{d}$}
  \psfrag{s}[c][c][1.0]{$\mathbf{k}_{a}$}
  \psfrag{t}[c][c][1.0]{$\mathbf{k}_{b}$}
  \psfrag{u}[c][c][1.0]{$\mathbf{k}_{c}$}
  \psfrag{v}[c][c][1.0]{$\mathbf{k}_{d}$}
  \psfrag{f}[c][c][1.0]{$f_\text{prism}$}
  \includegraphics[width=0.45\columnwidth]{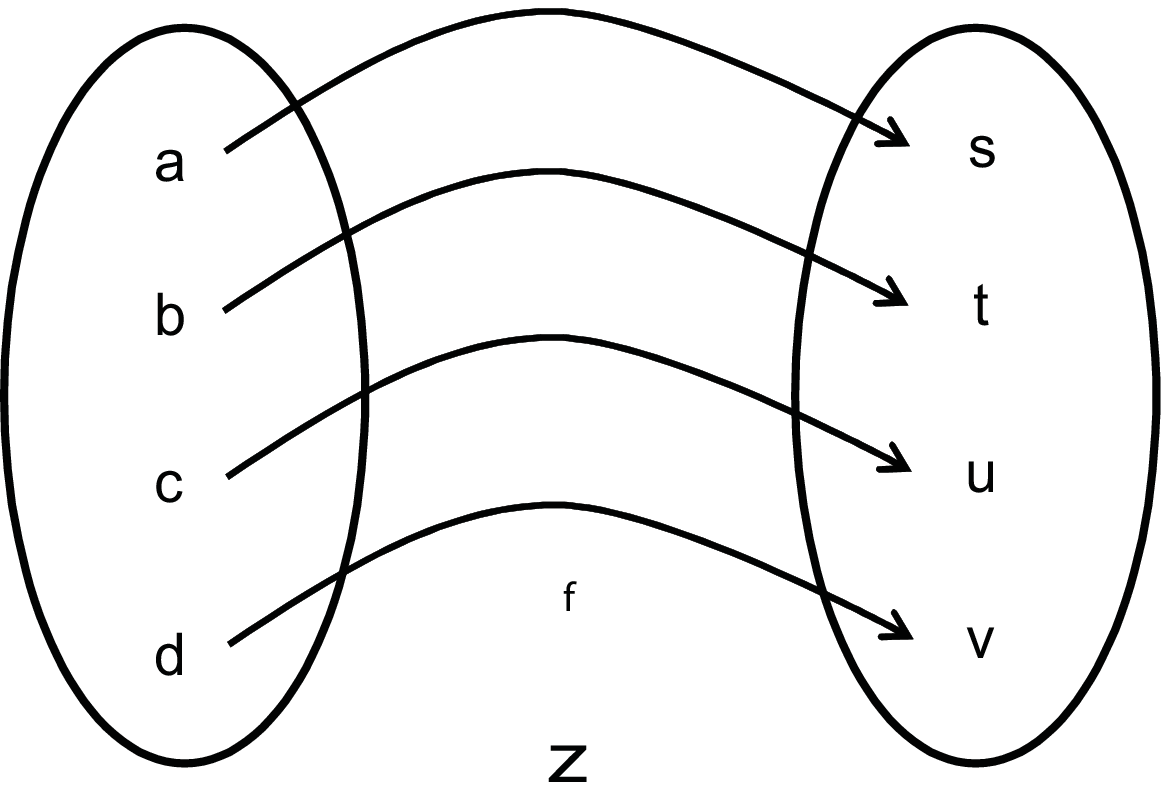}
  }
  \subfigure{
  \psfrag{z}[c][c][1.0]{(d)}
  \psfrag{s}[c][c][1.0]{$\omega_{a}$}
  \psfrag{t}[c][c][1.0]{$\omega_{b}$}
  \psfrag{u}[c][c][1.0]{$\omega_{c}$}
  \psfrag{v}[c][c][1.0]{$\omega_{d}$}
  \psfrag{a}[c][c][1.0]{$\mathbf{k}_{a}$}
  \psfrag{b}[c][c][1.0]{$\mathbf{k}_{b}$}
  \psfrag{c}[c][c][1.0]{$\mathbf{k}_{c}$}
  \psfrag{d}[c][c][1.0]{$\mathbf{k}_{d}$}
  \psfrag{f}[c][c][1.0]{$f_\text{prism}^{-1}$}
  \includegraphics[width=0.45\columnwidth]{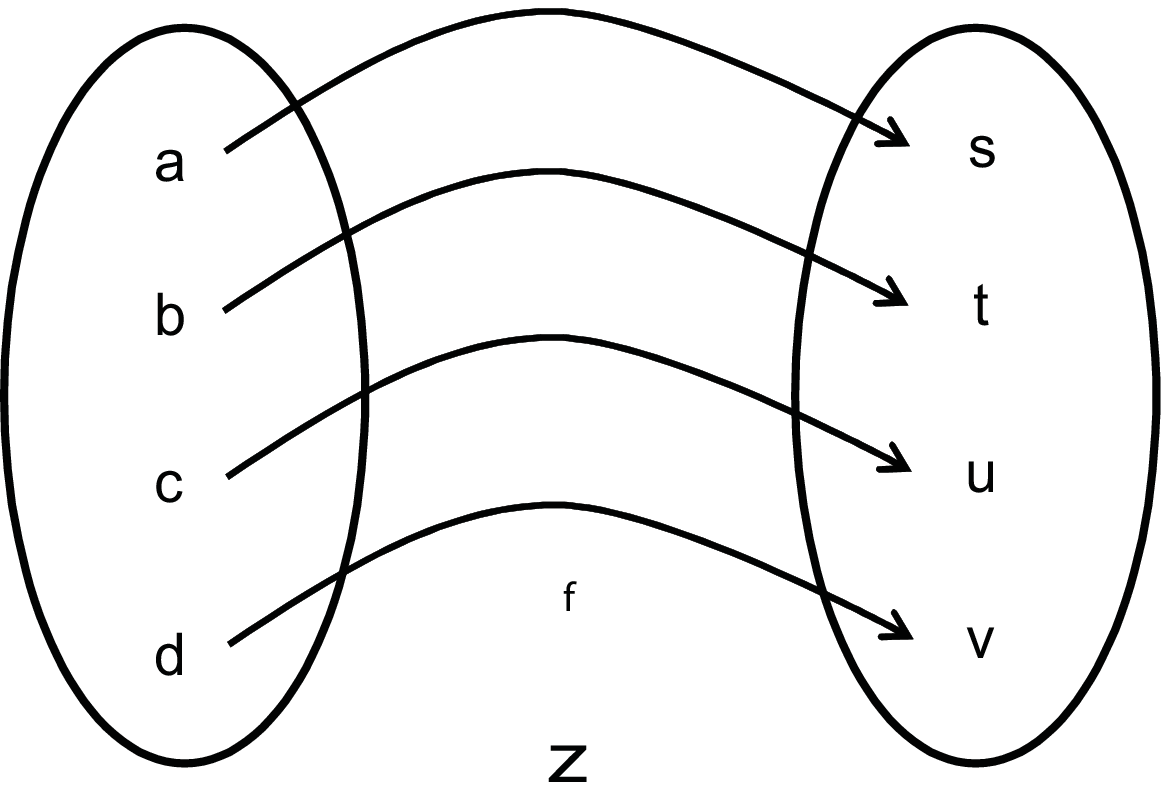}
  }
  \caption{Comparison of the conventional prism and the inverse prism, formed by the refractive indices $n_1$ and $n_2$ with $n_2>n_1$. (a)~Conventional prism, decomposing white light into its constituent colors, using spatial discontinuity [$n=n(\ve{r})$] and temporal dispersion [$n_2=n_2(\omega)$]. (b)~Corresponding mapping from temporal frequencies ($\omega$) to spatial frequencies ($\ve{k}$). (c)~Inverse prism, transforming
multidirectional light into direction-dependent new colors, using temporal discontinuity [$n=n(t)$] and spatial dispersion [$n_2=n_2(\ve{k})$]. (d)~Corresponding mapping from spatial to temporal frequencies.}
\label{fig:prism_comp}
\end{figure}

An optical prism, represented in Fig.~\ref{fig:prism_comp}(a), is a transparent device, which decomposes incident white light into its constituent colors by refracting them into different directions. This phenomenon was previously believed to result from the production of colors by the prism, and it is Newton who first found out, experimentally, that it was rather due to the decomposition of the colors already present in the incoming light. Newton later explained in his book Opticks \cite{newton1721opticks} that spectral decomposition itself results from the frequency dispersive nature of the glass medium forming the prism, whereby the refractive index is a function of frequency. From a mathematical perspective, one may consider a prism as a device that maps temporal frequencies ($\omega$) into spatial frequencies ($\ve{k}$), as shown in Fig.~\ref{fig:prism_comp}(b).

In this paper, we raise the question as to whether it may be possible to accomplish the inverse operation to that of the prism, namely mapping spatial frequencies into temporal frequencies, as mathematically suggested in Fig.~\ref{fig:prism_comp}(d). What would that physically mean? Reading out Fig.~\ref{fig:prism_comp}(d) from a physical perspective leads to the physical operation depicted in Fig.~\ref{fig:prism_comp}(c): waves with different directions but identical frequency would transform into waves with different frequencies without change of direction. A device performing this operation would naturally be referred to as an \emph{inverse prism}.

How could such an inverse prism be practically realized? Comparing Figs.~\ref{fig:prism_comp}(b) and \ref{fig:prism_comp}(d), as well as Figs.~\ref{fig:prism_comp}(a) and \ref{fig:prism_comp}(c), reveals that the inverse operation results from swapping space and time. The conventional and inverse prisms are thus the \emph{space-time dual} of each other. This consideration should naturally lead us to the realization of the inverse prism.

A conventional prism is characterized by a)~broken spatial symmetry [$n=n(\ve{r})$], which results in alteration (non-conservation) of the momentum of light, leading to spatial refraction, and b)~frequency dispersion [$n_2=n_2(\omega)$], typically present in glass materials at optical frequencies, to refract different frequencies into different directions. Therefore, the inverse inverse prism should be characterized by a$'$)~broken temporal symmetry [$n_2=n_2(t)$], resulting in alteration (non-conservation) of the energy, or frequency of light and \emph{temporal refraction}~\cite{mendoncca2000quantum,mendoncca2000theory}, and b$'$)~spatial dispersion [$n_2=n_2(\ve{k})$], for which a proper mechanism is to be devised to ``refract'' different directions (spatial frequencies) into different (temporal) frequencies. Thus, these dual properties constitute the key requirements to realize an inverse prism.

The first requirement, broken temporal symmetry [a$'$) above], can be accomplished by incorporating instantaneous (step-like) \emph{time switching} between two different media. Such temporal switching may be implemented by ionizing plasmas~\cite{kalluri2016electromagnetics,shvartsburg2005optics},
exciting nonlinearities~\cite{kaup1979space}, producing shock waves~\cite{reed2003color,reed2003reversed} or modulating varactors~\cite{taravati2017mixer}. The electrodynamics of wave propagation and light-matter interaction in temporal media and, in general, switched media have been well studied in past decades. In the late 1960ies, Morgenthaler solved the problem of light propagation in dielectric media with time varying wavenumbers, with specific application to step and sinusoidal temporal modulations~\cite{morgenthaler1958velocity}. About a decade later, his work was extended to time-varying media that are dispersive and include electromagnetic sources by Felsen and Whitman~\cite{felsen1970wave}, and to scattering by a spatial interface between vacuum and time-varying dielectric or dispersive media by Fante~\cite{fante1971wave}. From the 1970ies to the turn of the 20$^\text{th}$ century, studies of switched media were mostly restricted to plasma physics~\cite{kalluri2016electromagnetics,shvartsburg2005optics} and nonlinear optics~\cite{kaup1979space}. However, there has recently been a general regain of interest in the electrodynamics of light-matter interactions, in the context of temporal resonators and waveguides~\cite{xiao2011optical,xiao2011spectral,xiao2014reflection,plansinis2015temporal}, photonic crystals with spatiotemporal defects~\cite{yanik2004time,reed2003color,reed2003reversed}, space-time modulated graphene sheets~\cite{chamanara2016graphene}, and spatiotemporal metasurfaces~\cite{hadad2015space}.

The second requirement for realizing an inverse prism, spatial dispersion [b$'$) above], can be fulfilled by performing the aforementioned time switching from a standard isotropic medium ($n_1$) to an \emph{anisotropic} medium ($\te{n}_2$), which may alternatively be seen as the time-varying medium
\begin{equation}\label{eq:inv_prism_medium}
n(t)=n_{1}u(t_{0}-t)+\bar{\bar{n}}_{2}u(t-t_{0}),
\end{equation}
where $u(\cdot)$ is the step function and $t_0$ is the switching time. We shall consider here the typical optical case of non-magnetic materials, where $\mu_1=\mu_2=\mu_0$ and $n_1=\sqrt{\epsilon_1/\epsilon_0}$, $n_{2,ij}=\sqrt{\epsilon_{2,ij}/\epsilon_0}$ with $i,j=\{1,2,3\}$.

\begin{figure}
   \centering
   \subfigure{
   \psfrag{a}[c][c][0.8]{$ct$}
   \psfrag{b}[c][c][0.8]{$ct_{0}$}
   \psfrag{m}[c][c][0.8]{$n=\bar{\bar{n}}_{2}=\begin{pmatrix}
                                            n_{\|} & 0 & 0 \\
                                            0 & n_{\|} & 0 \\
                                            0 & 0 & n_{\bot}
                                          \end{pmatrix}$}
   \psfrag{f}[c][c][0.8]{$\theta$}
   \psfrag{h}[c][c][0.8]{$k_{x}$}
   \psfrag{g}[c][c][0.8]{$k_{z}$}
   \psfrag{k}[c][c][0.8]{$x$}
   \psfrag{d}[c][c][0.8]{$z$}
   \psfrag{s}[c][c][0.8]{s}
   \psfrag{p}[c][c][0.8]{$p$}
   \psfrag{e}[c][c][0.8]{$\omega_{1}$}
   \psfrag{n}[c][c][0.8]{$n=n_{1}$}
   \psfrag{i}[c][c][0.8]{$k_1=\frac{n_{1}\omega_{1}}{c}$}
   \psfrag{c}[c][c][0.8]{$x,y,z$}
   \psfrag{y}[c][c][0.8]{$y$}
   \psfrag{z}[c][c][1.0]{(a)}
   \includegraphics[width=0.5\columnwidth]{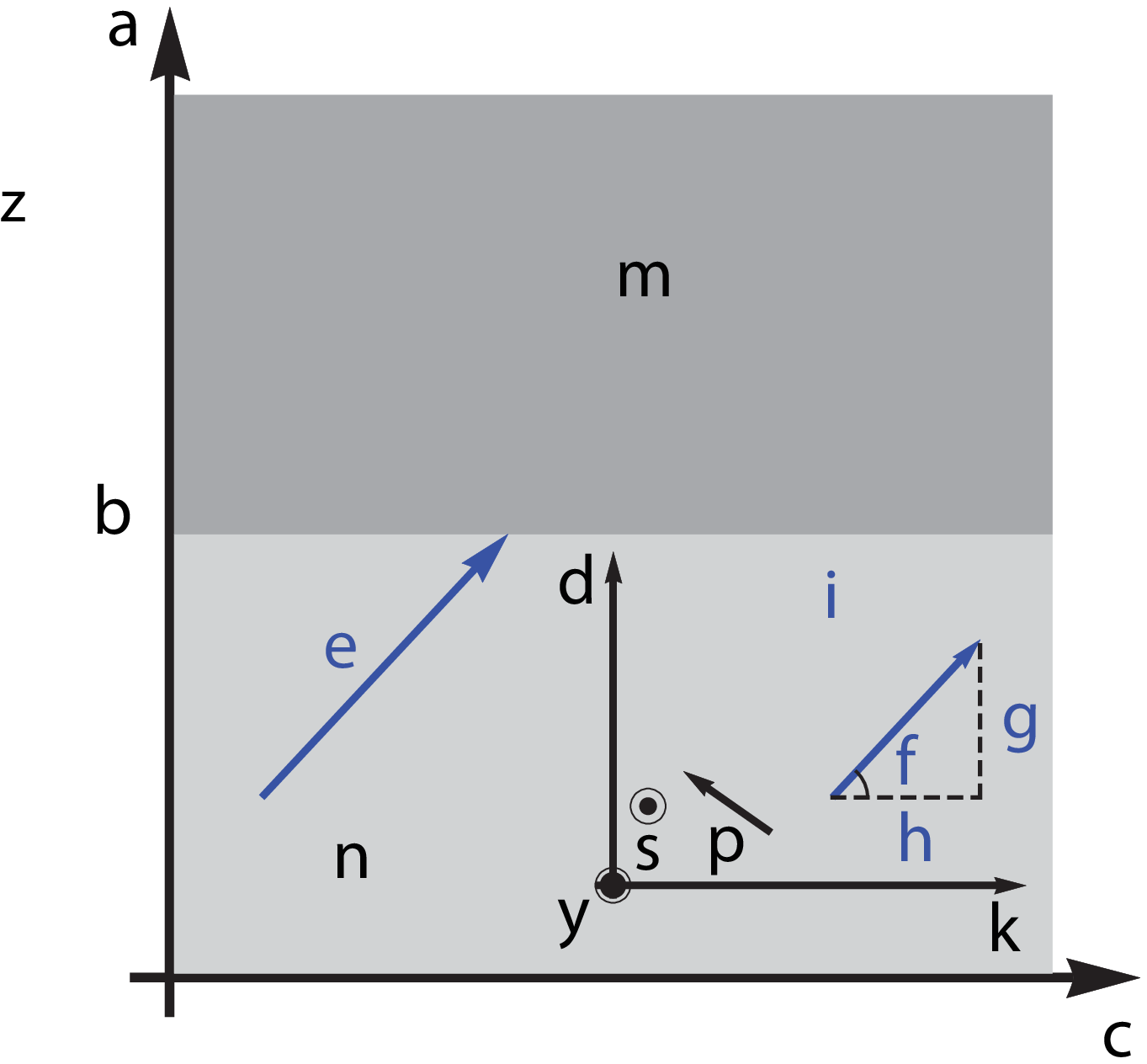}
   }
   \subfigure{
   \psfrag{a}[c][c][0.8]{$ck_{z}/n_{1}$}
   \psfrag{b}[c][c][0.8]{$ck_{x}/n_{1}$}
   \psfrag{c}[c][c][0.8]{$\omega_{1}$}
   \psfrag{d}[c][c][0.8]{$\omega_{s}$}
   \psfrag{e}[c][c][0.8]{$\omega_{p}$}
   \psfrag{z}[c][c][0.8]{$\omega_{\text{min}}$}
   \psfrag{y}[c][c][0.8]{$\omega_{\text{max}}$}
   \psfrag{f}[c][c][1.0]{(b)}
   \includegraphics[width=0.43\columnwidth]{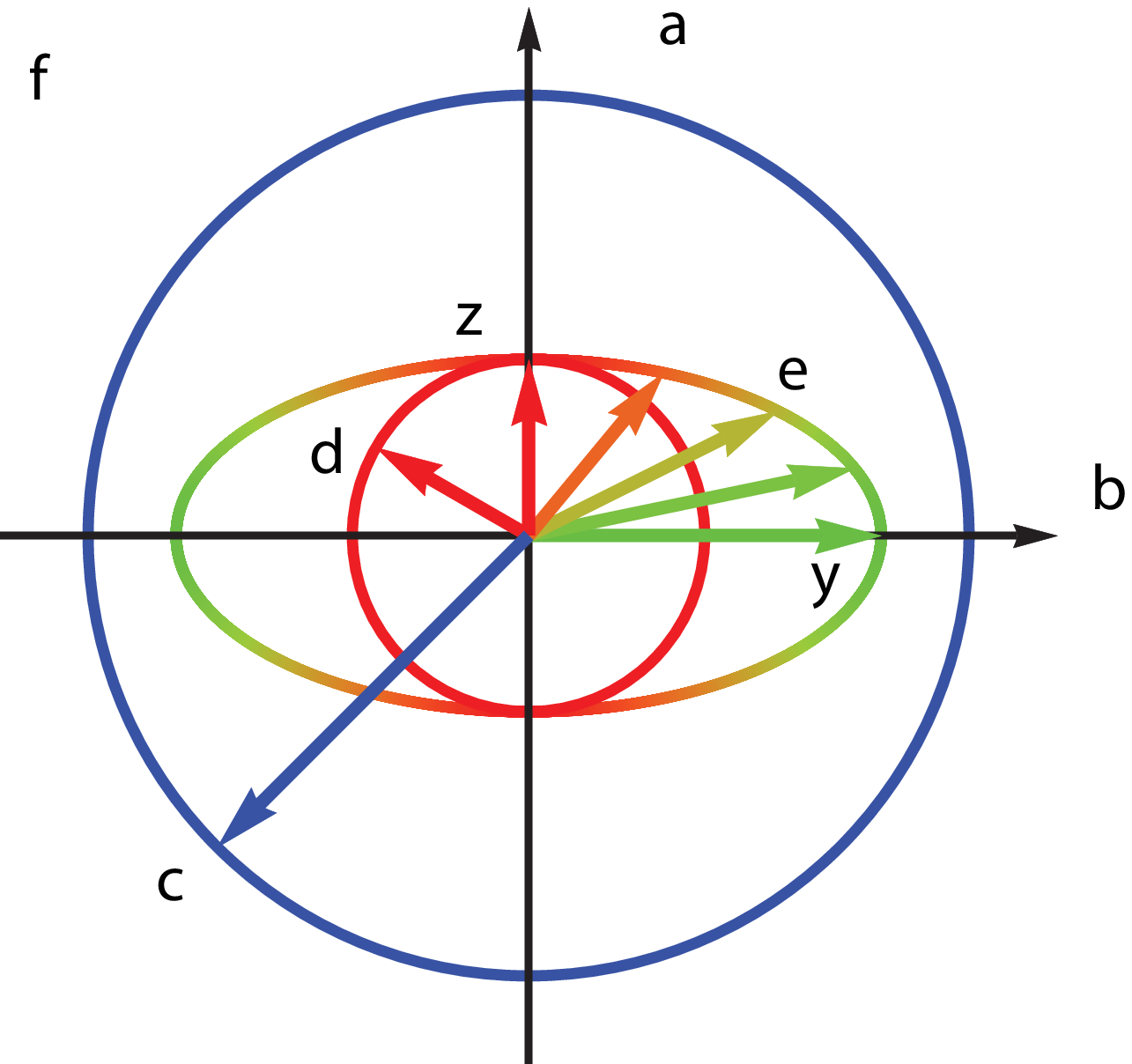}
   }\caption{Implementation of the inverse prism in Figure~\ref{fig:prism_comp}(c) and~(d), as the time-varying medium described by~\eqref{eq:inv_prism_medium}. (a)~Space-time representation of the inverse prism, whose anisotropy, here uniaxial, induces birefringence, with inset showing the $xz$ space, the $\text{s},\text{p}$ polarizations and the angle of wave propagation~$\theta$. (b)~Circular (isotropic) and elliptical (anisotropic) isofrequency curves corresponding to~\eqref{eq:omega_s} and~\eqref{eq:omega_p}, respectively, for $n_1<n_\bot<n_\|$.}\label{fig:Implementation_Inv_Prism}
\end{figure}

The anisotropic medium $\te{n}_2$, which may be generally characterized by the constitutive susceptibility tensor $\te{\chi}=\chi_{ij}$, is intrinsically spatially dispersive since its dispersion relation, $\omega=g\left(\chi_{ij},\mathbf{k}\right)$~\cite{born2013principles,kong1975theory,silveirinha2007metamaterial}, can be alternatively written as $\omega=ck_{0}/n\left(\mathbf{k}\right)$ with $n\left(\mathbf{k}\right)=ck_0/g\left(\chi_{ij},\mathbf{k}\right)$ and $k_0=\omega/c$ being the vacuum wavenumber associated with the vacuum speed $c$. In this paper, for simplicity but without loss of generality, we consider the case of uniaxial anisotropy, with the optical axis parallel to the $z$ axis and the refractive index tensor $\bar{\bar{n}}=\mathrm{diag}\{n_{\|},n_{\|},n_{\bot}\}$, as indicated in Fig.~ \ref{fig:Implementation_Inv_Prism}(a). This anisotropy allows one to decompose the problem into s (out-of-plane or $TE$ or ordinary) polarization and p (in-plane or $TM$ or extraordinary) polarization, as shown in the inset of the same figure.

Note that, the switching from medium $n_1$ to medium $\te{n}_2$ in~\eqref{eq:inv_prism_medium}, being purely temporal, incurs no change in spatial frequencies, i.e. $\ve{k}_2=\ve{k}_1=\ve{k}$: it is the momentum ($\ve{k}$) that is conserved in a temporal discontinuity problem~\cite{plansinis2015temporal,biancalana2007dynamics,deck2017wave}. This contrasts with the conservation of frequency ($\omega$) or energy ($\hbar\omega$) in a spatial discontinuity problem. Thus, the angle of propagation is not affected by the discontinuity, i.e. $\theta_2=\theta_1=\theta$.

Let us consider a harmonic incident plane wave with angular frequency $\omega_1$ propagating in the $x-z$ plane of the earlier isotropic medium ($n=n_{1}$, $t<t_0$) under the angle $\theta$ with respect to the $x$ axis, as illustrated in Fig.~\ref{fig:Implementation_Inv_Prism}(a). The corresponding dispersion relation may be written as
\begin{equation}\label{eq:k1}
k_1=n_1\frac{\omega_1}{c}=\sqrt{k_{x}^2+k_{z}^2}
\quad\text{or}\quad
\omega_1=\frac{c}{n_1}\sqrt{k_{x}^2+k_{z}^2},
\end{equation}
where $k_{x}$ and $k_{z}$ are the $x$ and $z$ components of wave vector $\ve{k}$, respectively. This relation describes a circle with radius $\omega_1$ in the isofrequency diagram normalized to $n_1/c$, as plotted in blue in Fig.~\ref{fig:Implementation_Inv_Prism}(b).

In the s-polarization problem, with electric field directed along the $y$~direction, the later medium is seen as purely isotropic, with refractive index $n_2=\sqrt{\epsilon_\|/\epsilon_0}=n_\|$. Therefore, accounting also for momentum conservation, the later dispersion relation is identical to~\eqref{eq:k1} with $n_1$ replaced by $n_\|$, i.e.
\begin{equation}\label{eq:omega_s}
k_\text{s}=n_\|\frac{\omega_\text{s}}{c}=\pm\sqrt{k_{x}^2+k_{z}^2}
\;\;\text{or}\;\;
\omega_\text{s}=\pm\frac{c}{n_\|}\sqrt{k_{x}^2+k_{z}^2},
\end{equation}
corresponding to the smaller circle in the isofrequency diagram of Fig.~\ref{fig:Implementation_Inv_Prism}(b), where the $+$ and $-$ denote the transmitted (forward) and reflected (backward) wave in the later medium, with respective scattering coefficients~\cite{biancalana2007dynamics,deck2017wave,supp_material_1} 
\begin{equation}
T_\text{s}=\frac{n_1\left(n_1+n_\|\right)}{2n_\|^2},
\label{eq:Transmission_s}
\end{equation}
\begin{equation}
R_\text{s}=\frac{n_1\left(n_1-n_\|\right)}{2n_\|^2}.
\label{eq:Reflection_s}
\end{equation}
The p-polarization problem, which is of greater interest, is more complex. Its dispersion relation is~\cite{supp_material_2}
\begin{subequations}\label{eq:omega_p}
\begin{equation}\label{eq:omega_p_k}
n_\|^2n_\bot^2k_0^2
=n_\|^2n_\bot^2\left(\frac{\omega_\text{p}}{c}\right)^2
=n_\|^2k_x^2+n_\bot^2k_z^2,
\end{equation}
which becomes, upon dividing by $n_\|^2n_\bot^2/c^2$ and taking the square root,
\begin{equation}\label{eq:omega_p_w}
 \omega_\text{p}
=\omega_{\bot}=\pm\sqrt{\left(\frac{c}{n_{\bot}}\right)^{2}k_{x}^{2}
+\left(\frac{c}{n_{\|}}\right)^{2}k_{z}^{2}},
\end{equation}
\end{subequations}
where the $+$ and $-$ denote again the transmitted (forward) and reflected (backward) wave in the later medium with respective scattering coefficients~\cite{supp_material_1}
\begin{equation}
T_\text{p}=\frac{n_{1}^2\sqrt{n_\bot^4k_z^2+n_\|^4k_x^2}}{n_\|^2n_\bot^2\sqrt{k_x^2+k_z^2}}\left(\frac{\omega_1+\omega_{p}}{2\omega_{1}}\right),
\label{eq:Transmission_p}
\end{equation}
\begin{equation}
R_\text{p}=\frac{n_{1}^2\sqrt{n_\bot^4k_z^2+n_\|^4k_x^2}}{n_\|^2n_\bot^2\sqrt{k_x^2+k_z^2}}\left(\frac{\omega_1-\omega_{p}}{2\omega_{1}}\right).
\label{eq:Reflection_p}
\end{equation}

The relation~\eqref{eq:omega_p_w} describes an ellipse with major and minor semi-axes $\omega_\text{max}=(n_1/n_\bot)\omega_1$ and $\omega_\text{min}=(n_1/n_\|)\omega_1=\omega_\text{s}$, respectively, as plotted in gradient color in Fig.~\ref{fig:Implementation_Inv_Prism}(b). Thus, the scattered frequency is a function of the wave direction, or $\omega=\omega(\ve{k})$, which corresponds to the sought after inverse prism operation [Fig.~\ref{fig:prism_comp}(c) and (d)].

\begin{figure}
  \centering
  \subfigure{
  \psfrag{b}[c][c][0.9]{$z$}
  \psfrag{c}[c][c][0.9]{$x$}
  \psfrag{f}[c][c][0.9]{$t=t_{0}$}
  \psfrag{d}[c][c][0.9][60]{{$\omega_{1}$}}
  \psfrag{e}[c][c][0.9][60]{$\omega_{\text{p}1}$}
  \psfrag{h}[c][c][0.9][60]{$\omega_{\text{p}2}$}
  \psfrag{j}[c][c][0.9][60]{$\omega_{\text{p}3}$}
  \psfrag{g}[c][c][0.9][60]{$\omega_{\text{s}}$}
  \includegraphics[width=0.7\columnwidth]{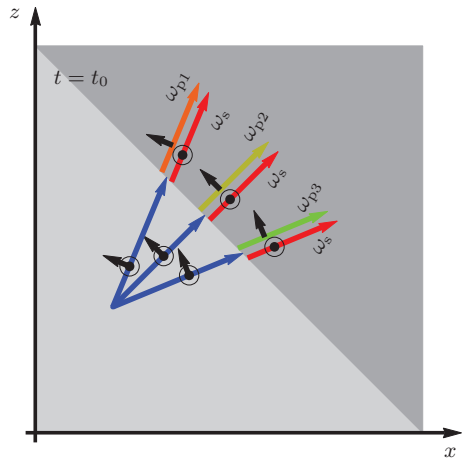}
  }
\caption{Temporal frequency birefringence of the uniaxial inverse prism of Fig.~\ref{fig:Implementation_Inv_Prism}. In a hybrid wave, the s-polarized and p-polarized parts experience ordinary (monochromatic) and extraordinary (polychromatic) temporal refraction, respectively.}
  \label{fig:Frequency_Birefringence}
\end{figure}

We have just considered the s-polarization and p-polarization problems separately. However, a general wave, as for instance unpolarized light, may carry both polarizations. Such a hybrid wave would be split by the anisotropic inverse prism of Fig.~\ref{fig:Implementation_Inv_Prism} into an s-wave and a p-wave seeing the frequency transformations~\eqref{eq:omega_s} and $\eqref{eq:omega_p}$, respectively. The prism thus exhibits temporal frequency  birefringence, specifically \emph{chromatic refraction birefringence}, which is a type of birefringence not found in structures with spatial discontinuities or isotropic temporal discontinuities. This birefringence phenomenon is illustrated in Fig.~\ref{fig:Frequency_Birefringence}. Monochromatic light rays with identical frequency, $\omega_{1}$, propagating in different directions in the earlier medium are scattered without change of direction as new monochromatic and polychromatic waves for the s-polarization and p-polarization cases, respectively.

Another interesting perspective of the inverse prism is that it bears \emph{analogy with Snell law} of refraction at a spatial discontinuity. For the case of s-polarization in the uniaxial system of Fig.~\ref{fig:Implementation_Inv_Prism},  taking the ratio of \eqref{eq:omega_s} and~\eqref{eq:k1} yields
\begin{equation}\label{eq:Snell_s}
  n_1\omega_1=\pm n_{\|}\omega_{\text{s}}.
\end{equation}
This relation, illustrated in Fig.~\ref{fig:Frequency_Birefringence}, may be considered as the \emph{dual of Snell law} between isotropic media ($n_1\sin\theta_1=n_2\sin\theta_2$): it is a transformation of temporal frequency ($\omega_1$ to $\omega_2=\omega_\text{s}$), instead of spatial frequency or angle ($\theta_1$ to $\theta_2$), induced by a temporal interface, instead of a spatial interface, between two distinct media ($n_1$ and $n_2=n_\|$). This corresponds to the conventional time refraction discussed in~\cite{mendoncca2000quantum,mendoncca2000theory}, and we therefore refer to it here as the \emph{ordinary} temporal refraction law.

In the case of p-polarization, the temporal refraction relation, obtained by taking the ratio of \eqref{eq:omega_p_w} to \eqref{eq:k1}, enforcing $\mathbf{k}_1=\mathbf{k}_2=\mathbf{k}$, and writing $k_x=|\ve{k}|\cos(\theta)$ and $k_z=|\ve{k}|\sin(\theta)$, takes the more complex form
\begin{equation}\label{eq:Snell_p}
  n_1\omega_1=\pm \frac{n_{\|}n_{\bot}}{\sqrt{n_{\|}^2\cos^2(\theta)+n_{\bot}^2\sin^2(\theta)}}\omega_\text{p}
  =n(\mathbf{k})\omega_\text{p},
\end{equation}
where $\theta$ is defined in the inset of Fig.~\ref{fig:Implementation_Inv_Prism}(a). This relation, which is also illustrated in Fig.~\ref{fig:Frequency_Birefringence}, is the \emph{extraordinary} counterpart of~\eqref{eq:Snell_s}, and may be considered as the \emph{dual of Snell law} with an anisotropic medium ($n_1\sin\theta_1=n_2(\ve{k})\sin\theta_2$).

\begin{figure}
\centering
\subfigure{
\psfrag{c}[c][c][1.0]{(a)}
\psfrag{b}[c][t][0.85]{$E_z$}
\psfrag{a}[l][c][0.85]{$E_y$}
\includegraphics[width=0.48\columnwidth]{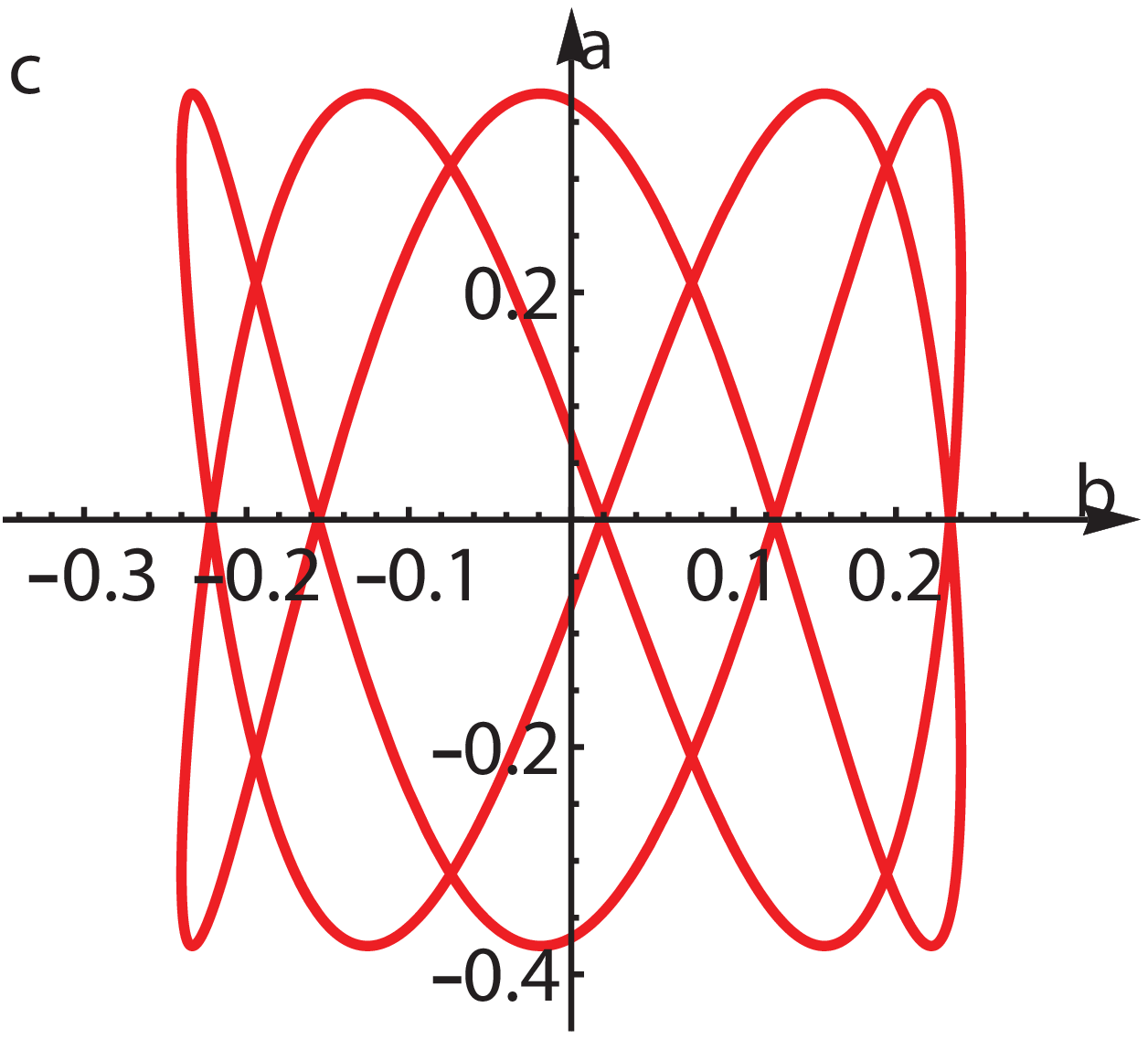}
}
\subfigure{
\psfrag{c}[c][c][1.0]{(b)}
\psfrag{a}[c][t][0.85]{$E_y$}
\psfrag{b}[l][c][0.85]{$E_z$}
\includegraphics[width=0.44\columnwidth]{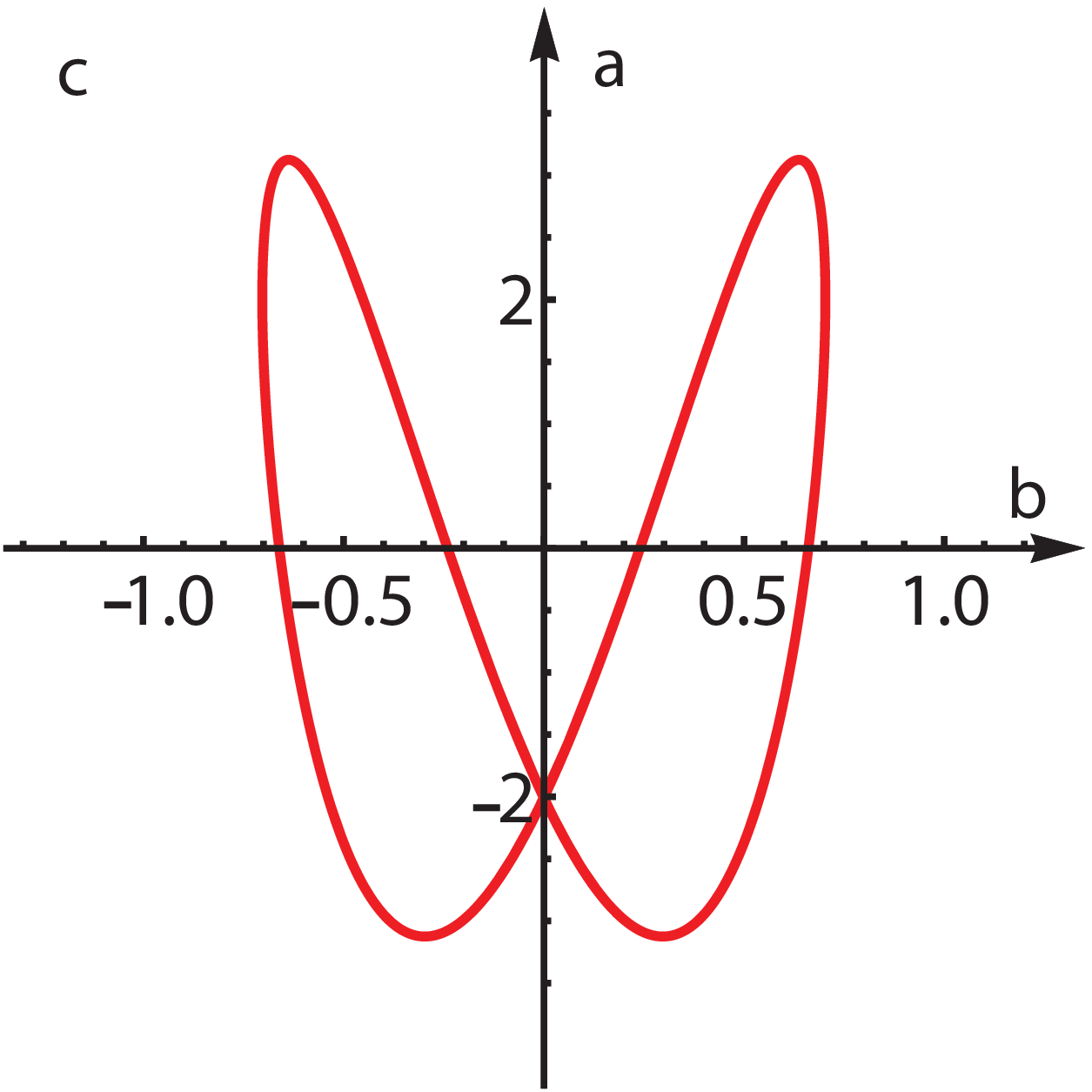}
}
\caption{Lissajous polarization cycle of the inverse prism of Fig.~\ref{fig:Implementation_Inv_Prism} for 2 different sets of parameters. (a)~$E_y=2$, $E_z=1$, $\omega_1/c=10$, $n_1=1$, $n_\|=5$, $n_\bot=2$, $\phi_{y}=-\pi/8$, and $\phi_{z}=0$. (b)~$E_y=10$, $E_z=20$, $\omega_1/c=16$, $n_1=1$, $n_\|=8$, $n_\bot=4$, $\phi_{y}=-\pi/18$, and $\phi_{z}=\pi/6$.}
\label{fig:Lissajous_Curves}
\end{figure}

Since it imparts different frequency jumps to s-polarized and p-polarized waves propagating in a given direction, as illustrated Fig.~\ref{fig:Frequency_Birefringence}, the uniaxial inverse prism may be expected to induce unusual polarization transformation in the case of a hybrid (mixed s- and p-polarized) wave. To investigate this, let us consider a hybrid plane wave that is linearly polarized in the $y-z$ plane and propagating along the $x$-axis in the earlier isotropic medium [see inset of Fig.~\ref{fig:Implementation_Inv_Prism}(a)]. The electric field of such a wave may be written
\begin{equation}
\mathbf{E}_{1}(x,t)=\mathbf{e}_{y}E_{y}\cos(k_{x}x-\omega_{1}t)+\mathbf{e}_{z}E_{z}\cos(k_{x}x-\omega_{1}t).
\label{eq:Lissajous_a}
\end{equation}
At the temporal interface, the $y$ (s-polarized) component of the electric field undergoes the frequency shift $\omega_{1}$ to $\omega_\text{s}$ (ordinary wave), while the $z$ (p-polarized) component undergoes the frequency shift $\omega_{1}$ to $\omega_\text{p}$ (extraordinary wave). Moreover, as previously mentioned, each of the s and p component splits into transmitted and reflect waves with respective coefficients $T_\text{p,s}$ and $R_\text{p,s}$ provided in Eqs.~\eqref{eq:Transmission_s} to~\eqref{eq:Reflection_p}. Therefore, the wave transmitted ($+$) at the interface reads
\begin{equation}
\begin{split}
\mathbf{E}_{2}^+(x,t)=&\mathbf{e}_{y}T_\text{s}E_{y}\cos[k_{x}x-\omega_\text{s}(t-t_0)-\omega_1t_0]\\
&+\mathbf{e}_{z}T_\text{p}E_{z}\cos[k_{x}x-\omega_\text{p}(t-t_0)-\omega_1t_0],
\end{split}
\label{eq:Lissajous_b}
\end{equation}
which may be alternatively written as
\begin{subequations}\label{eq:Lissajous_2}
\begin{equation}
\mathbf{E}_{2}^+(x,t)=\mathbf{e}_{y}A\cos(\omega_\text{s}t+\phi_{y})+\mathbf{e}_{z}B\cos(\omega_\text{p}t+\phi_{z}),
\label{eq:Lissajous_2}
\end{equation}
with
\begin{equation}
A=T_\text{s}E_{y},\quad B=T_\text{p}E_{z},
\end{equation}
where $T_\text{s}$ and $T_\text{p}$ are available in~\eqref{eq:Transmission_s} and~\eqref{eq:Transmission_p}, respectively, and
\begin{equation}
\phi_{y}=-k_{x}x-(\omega_\text{s}-\omega_1)t_0,
\;
\phi_{z}=-k_{x}x-(\omega_\text{p}-\omega_1)t_0.
\end{equation}
\end{subequations}
Due to the different temporal variation rates in its $y$ component ($\omega_\text{s}$) and $z$ component ($\omega_\text{p}$), the field~\eqref{eq:Lissajous_2} has a polarization response that is more complex than the usual elliptical polarization. Indeed, Eq.~\eqref{eq:Lissajous_2} represents the parametric system, with parameter $t$,
\begin{equation}
\begin{aligned}
E_{2y}^+(t)&=A\cos(\omega_\text{s}t+\phi_{y}),\\
E_{2z}^+(t)&=B\cos(\omega_\text{p}t+\phi_{z}),
\end{aligned}
\label{eq:Lissajous_param}
\end{equation}
which describes Lissajous curves~\cite{surhone2010lissajous}. Thus, the tip of the electric field vector in~\eqref{eq:Lissajous_2} traces a Lissajous curve in space and time, and the inverse prism therefore transforms linear polarization into~\emph{Lissajous polarization}, represented in Fig.~\ref{fig:Lissajous_Curves} for three different media parameter sets. Lissajous polarization is naturally a generalization of elliptic and circular polarizations, which occur here in the limit case of a later isotropic medium for which $\omega_\text{s}=\omega_\text{p}$.

In this letter, we have introduced the concept of the inverse prim as the dual of the conventional prism. From this duality, we have deduced an implementation of the inverse prism based on temporal discontinuity and spatial dispersion provided by anisotropy. We have particularly considered the case of uniaxial anisotropy, for which we have pointed out the following unique properties of the inverse prism. First, it exhibits chromatic refraction birefringence. Second, it performs  ordinary (monochromatic) and extraordinary (polychromatic) temporal refraction for waves polarized perpendicularly to and in the plane of the optical axis, respectively. Third, it transforms linear polarization into Lissajous polarization. Thus, the inverse prism represents a new canonical concept in physics, which may lead to a diversity of novel optical devices, particularly for applications involving spin-orbit angular momentum manipulations.

\newpage

\appendix

\section{Derivation of the Transmission and Reflection Coefficients}

We first derive the transmission and reflection coefficients for p-polarization, and next deduce those for s-polarization from them.

In reference to Fig.~2(a) of the main text, the electric and magnetic fields of the p-polarized plane wave
\begin{equation}\label{eq:E_1_p_pol}
  \ve{E}_1=\frac{1}{\omega_1\epsilon_0}\left(\ve{e}_x\frac{1}{n_1^2}k_z-\ve{e}_z\frac{1}{n_1^2}k_x\right)e^{i\left(k_xx+k_zz-\omega_1t\right)},
\end{equation}
\begin{equation}\label{eq:H_1_p_pol}
  \ve{H}_1=\ve{e}_y e^{i\left(k_xx+k_zz-\omega_1t\right)},
\end{equation}
splits at $t=t_0$ into a forward transmitted (forward) and a reflected (backward) wave with the angular frequency $\omega_\text{p}$ as

\begin{widetext}
\begin{equation}\label{eq:E_2_p_pol}
  \ve{E}_2=\ve{E}^++\ve{E}^-=\frac{1}{\omega_\text{p}\epsilon_0}\left(\ve{e}_x\frac{1}{n_\|^2}k_z-\ve{e}_z\frac{1}{n_\bot^2}k_x\right)
  e^{i\left(k_xx+k_zz\right)}\left(Ae^{-i\omega_\text{p}(t-t_0)}-Be^{+i\omega_\text{p}(t-t_0)}\right),
\end{equation}
\begin{equation}\label{eq:H_2_p_pol}
  \ve{H}_2=\ve{H}^++\ve{H}^-=\ve{e}_ye^{i\left(k_xx+k_zz\right)}\left(Ae^{-i\omega_\text{p}(t-t_0)}+Be^{+i\omega_\text{p}(t-t_0)}\right),
\end{equation}
\end{widetext}

Since the discontinuity is purely temporal, the total charge is conserved and therefore the total fields $\ve{D}$ and $\ve{H}$ are conserved according to the Maxwell-Gauss equations at $t=t_0$. Therefore, we have
\begin{equation}\label{eq:D2_eq_D1_p_pol}
  \frac{1}{\omega_\text{p}}\left(A-B\right)=\frac{1}{\omega_1}e^{-i\omega_1t_0},
\end{equation}
\begin{equation}\label{eq:H2_eq_H1_p_pol}
  \left(A+B\right)=e^{-i\omega_1t_0},
\end{equation}
which gives
\begin{equation}\label{eq:A_p_pol}
  A=\frac{\omega_1+\omega_\text{p}}{2\omega_1}e^{-i\omega_1t_0},
\end{equation}
\begin{equation}\label{eq:B_p_pol}
  B=\frac{\omega_1-\omega_\text{p}}{2\omega_1}e^{-i\omega_1t_0}.
\end{equation}
Inserting~\eqref{eq:A_p_pol} and \eqref{eq:B_p_pol} into \eqref{eq:E_2_p_pol} and \eqref{eq:E_1_p_pol} yield the electric fields, from the
the transmission and reflection coefficients are obtained as
\begin{equation}\label{eq:Trans_p}
  T_\text{p}=\frac{E_2^+}{E_1}=\frac{n_{1}^2\sqrt{n_\bot^4k_z^2+n_\|^4k_x^2}}{n_\|^2n_\bot^2\sqrt{k_x^2+k_z^2}}\left(\frac{\omega_1+\omega_\text{p}}{2\omega_{1}}\right),
\end{equation}
\begin{equation}\label{eq:Ref_p}
  R_\text{p}=\frac{E_2^-}{E_1}=\frac{n_{1}^2\sqrt{n_\bot^4k_z^2+n_\|^4k_x^2}}{n_\|^2n_\bot^2\sqrt{k_x^2+k_z^2}}\left(\frac{\omega_1-\omega_\text{p}}{2\omega_{1}}\right),
\end{equation}
where $E_1$, $E_2^+$, and $E_2^-$ are the amplitudes of the $\ve{E_1}$, $\ve{E_2^+}$, and $\ve{E_2^-}$ vectors, respectively.

The transmission and reflection coefficients for the s-polarization case can be obtained through a similar method. However, as a simpler alternative, one can obtain the transmission and reflection coefficients of the s-polarization problem by simply replacing $n_\bot$ with $n_\|$ and $\omega_\text{p}$ with $\omega_\text{s}$, respectively, in \eqref{eq:Trans_p} and \eqref{eq:Ref_p}, which yields
\begin{equation}\label{eq:Trans_s_1}
  T_\text{s}=\frac{n_1^2}{n_\|^2}\left(\frac{\omega_1+\omega_\text{s}}{2\omega_{1}}\right),
\end{equation}
\begin{equation}\label{eq:Ref_s_1}
  R_\text{s}=\frac{n_1^2}{n_\|^2}\left(\frac{\omega_1-\omega_\text{s}}{2\omega_{1}}\right),
\end{equation}
which, upon using $\omega_\text{s}=\left(n_\|/n_1\right)\omega_1$, reads in terms of refractive indices
\begin{equation}\label{eq:Trans_s_2}
  T_\text{s}=\frac{n_1\left(n_1+n_\|\right)}{2n_\|^2},
\end{equation}
\begin{equation}\label{eq:Trans_s_2}
  R_\text{s}=\frac{n_1\left(n_1-n_\|\right)}{2n_\|^2}.
\end{equation}

\section{Dispersion Relation in a Uniaxial Dielectric Medium}

In a source-free dielectric uniaxial medium with permittivity tensor $\te{\epsilon}=\epsilon_0\te{\epsilon}_\text{r}=\epsilon_0\mathrm{diag}\left\{n_{\|}^2,n_{\|}^2,n_{\bot}^2\right\}$ and permeability $\mu=\mu_0$, where $\epsilon_0$ and $\mu_0$ are the vacuum permittivity and permeability, Maxwell equations read
\begin{equation}\label{eq:Curl_E_1}
  \ve{\nabla}\times\ve{E}=-\mu_0\frac{\partial{\ve{H}}}{\partial{t}},
\end{equation}
\begin{equation}\label{eq:Curl_H_1}
  \ve{\nabla}\times\ve{H}=+\frac{\partial{\ve{D}}}{\partial{t}},
\end{equation}
\begin{equation}\label{eq:Div_D_1}
  \ve{\nabla}\cdot\ve{D}=0,
\end{equation}
\begin{equation}\label{eq:Div_H_1}
  \ve{\nabla}\cdot\ve{H}=0,
\end{equation}
where $\ve{E}$, $\ve{H}$, and $\ve{D}$ are the electric, magnetic, and displacement fields, respectively. For a general plane wave propagating along the wave vector $\ve{k}=\ve{e}_xk_x+\ve{e}_yk_y+\ve{e}_zk_z$ and with a time harmonic dependence $\exp(-i\omega t)$, where $\omega$ is the angular frequency, Maxwell equations simplify to
\begin{equation}\label{eq:Curl_E_2}
  i\ve{k}\times\ve{E}=+i\omega\mu_0\ve{H},
\end{equation}
\begin{equation}\label{eq:Curl_H_2}
   i\ve{k}\times\ve{H}=-i\omega\ve{D},
\end{equation}
\begin{equation}\label{eq:Div_D_2}
  i\ve{k}\cdot\ve{D}=0,
\end{equation}
\begin{equation}\label{eq:Div_H_2}
  i\ve{k}\cdot\ve{H}=0,
\end{equation}
Cross multiplying \eqref{eq:Curl_E_2} by $\ve{k}$ yields
\begin{equation}\label{eq:Cross_1}
   i\ve{k}\times\left(\ve{k}\times\ve{E}\right)=+i\omega\mu_0\ve{k}\times\ve{H}.
\end{equation}
Using the identity $\ve{a}\times\left(\ve{b}\times\ve{c}\right)=\left(\ve{a}\cdot\ve{c}\right)\ve{b}-\left(\ve{a}\cdot\ve{b}\right)\ve{c}$, this relation becomes
\begin{equation}\label{eq:Cross_1_result}
   i\left(\ve{k}\cdot\ve{E}\right)\ve{k}-ik^2\ve{E}=+i\omega\mu_0\ve{k}\times\ve{H}
\end{equation}
Substituting $\ve{D}=\te{\epsilon}\cdot\ve{E}$ into~\eqref{eq:Curl_H_2} and \eqref{eq:Div_D_2}, and subsequently replacing $\ve{k}\times\ve{H}$ and $\ve{k}\cdot\ve{E}$ yields
\begin{equation}\label{eq:Simplified_1}
   \left(\ve{k}\cdot\ve{E}\right)\ve{k}-k^2\ve{E}+\frac{\omega^2}{c^2}\te{\epsilon}_r\cdot\ve{E}=0.
\end{equation}
Upon some algebraic manipulations, this equation can be recast in the matrix form
\begin{equation}\label{eq:Matrix_1}
  \te{M}\cdot\ve{E}=0,
\end{equation}
where

\begin{widetext}
\begin{equation}\label{eq:M_Matrix}
  \te{M}=\begin{pmatrix}
                                            -k_y^2-k_z^2+n_\|^2\left(\omega^2/c^2\right) & k_x k_y & k_x k_z \\
                                            k_x k_y & -k_x^2-k_z^2+n_\|^2\left(\omega^2/c^2\right) & k_y k_z \\
                                            k_x k_z & k_y k_z & -k_x^2-k_y^2+n_\bot^2\left(\omega^2/c^2\right)
                                          \end{pmatrix}.
\end{equation}
\end{widetext}

The nontrivial solution of this system is obtained by nullifying the determinant of $\te{M}$, which leads to

\begin{widetext}
\begin{equation}\label{eq:Det_M}
  \left[k_x^2+k_y^2+k_z^2-n_\|^2\left(\omega^2/c^2\right)\right]\left[n_\|^2k_x^2+n_\|^2k_y^2+n_\bot^2k_z^2-n_\|^2n_\bot^2\left(\omega^2/c^2\right)\right]=0,
\end{equation}
\end{widetext}

which is the dispersion relation of a general uniaxial dielectric medium.

Since we are interested in plane waves propagating along $x-z$ plane in the main text, we let $k_y=0$, and obtain
\begin{equation}\label{eq:Det_M_plane}
  \left[k_x^2+k_z^2-n_\|^2\left(\omega^2/c^2\right)\right]\left[n_\|^2k_x^2+n_\bot^2k_z^2-n_\|^2n_\bot^2\left(\omega^2/c^2\right)\right]=0.
\end{equation}
The left branch of~\eqref{eq:Det_M_plane} is the dispersion relation for s-polarization while the right branch is the dispersion relation for p-polarization.

\bibliography{ReferenceList_Inv_Prism_Paper}

\end{document}